\documentclass[amsmath,amssymb,superscriptaddress,preprint,floatfix,footinbib]{revtex4-1}
\usepackage[]{graphicx}
\usepackage{tabularx}
\usepackage[usenames,dvipsnames]{color}
\usepackage{soul}
\usepackage{bm}

\usepackage{times}
\usepackage{amsfonts}
\usepackage{mathrsfs}
\usepackage{graphicx}% Include figure files
\usepackage{dcolumn}% Align table columns on decimal point
\usepackage{bm}% bold math
\usepackage{color}
\usepackage[colorlinks,bookmarks=false,citecolor=blue,linkcolor=red,urlcolor=blue]{hyperref}

\usepackage{multirow}
\usepackage{physics}

\begin{document}

\title{The electronic structure of intertwined kagome, honeycomb, and triangular sublattices of the intermetallics MCo$_2$Al$_9$}

\author{Chiara Bigi}\email{chiara.bigi@synchrotron-soleil.fr}
\affiliation{SUPA, School of Physics and Astronomy, University of St Andrews, St Andrews KY16 9SS, UK}\affiliation{Synchrotron SOLEIL, F-91190 Saint-Aubin, France}
\author{Sahar Pakdel}
\affiliation{CAMD, Department of Physics, Technical University of Denmark, 2800 Kgs. Lyngby, Denmark}
\author{Micha{\l} J. Winiarski}
\affiliation{Faculty of Applied Physics and Mathematics, Advanced Materials Centre, Gdansk University of Technology, Narutowicza 11/12, 80-233, Gdansk, Poland}
\author{Pasquale Orgiani}
\affiliation{Istituto Officina dei Materiali (IOM)-CNR, Laboratorio TASC, Area Science Park, S.S.14, Km 163.5, 34149 Trieste, Italy}
\author{Ivana Vobornik}
\affiliation{Istituto Officina dei Materiali (IOM)-CNR, Laboratorio TASC, Area Science Park, S.S.14, Km 163.5, 34149 Trieste, Italy}
\author{Jun Fujii}
\affiliation{Istituto Officina dei Materiali (IOM)-CNR, Laboratorio TASC, Area Science Park, S.S.14, Km 163.5, 34149 Trieste, Italy}
\author{Giorgio Rossi}
\affiliation{Istituto Officina dei Materiali (IOM)-CNR, Laboratorio TASC, Area Science Park, S.S.14, Km 163.5, 34149 Trieste, Italy}\affiliation{Department of Physics, University of Milano, 20133 Milano, Italy}
\author{Vincent Polewczyk}
\affiliation{Istituto Officina dei Materiali (IOM)-CNR, Laboratorio TASC, Area Science Park, S.S.14, Km 163.5, 34149 Trieste, Italy}
\author{Phil D.C. King}
\affiliation{SUPA, School of Physics and Astronomy, University of St Andrews, St Andrews KY16 9SS, UK}
\author{Giancarlo Panaccione}
\affiliation{Istituto Officina dei Materiali (IOM)-CNR, Laboratorio TASC, Area Science Park, S.S.14, Km 163.5, 34149 Trieste, Italy}
\author{Tomasz Klimczuk} 
\affiliation{Faculty of Applied Physics and Mathematics, Advanced Materials Centre, Gdansk University of Technology, Narutowicza 11/12, 80-233, Gdansk, Poland}
\author{Kristian Sommer Thygesen}\email{thygesen@fysik.dtu.dk}
\affiliation{CAMD, Department of Physics, Technical University of Denmark, 2800 Kgs. Lyngby, Denmark}
\author{Federico Mazzola}\email{federico.mazzola@unive.it}
\affiliation{Department of Molecular Sciences and Nanosystems, Ca’ Foscari University of Venice, 30172 Venice, Italy}\affiliation{Istituto Officina dei Materiali (IOM)-CNR, Laboratorio TASC, Area Science Park, S.S.14, Km 163.5, 34149 Trieste, Italy}

\date{\today}

\begin{abstract}
Intermetallics are an important playground to stabilize a large variety of physical phenomena, arising from their complex crystal structure. The ease of their chemical tuneabilty makes them suitable platforms to realize targeted electronic properties starting from the symmetries hidden in their unit cell. Here, we investigate the family of the recently discovered intermetallics MCo$_2$Al$_9$ (M: Sr, Ba) and we unveil their electronic structure for the first time. By using angle-resolved photoelectron spectroscopy and density functional theory calculations, we discover the existence of Dirac-like dispersions as ubiquitous features in this family, coming from the hidden kagome and honeycomb symmetries embedded in the unit cell. Finally, from calculations, we expect that the spin-orbit coupling is responsible for opening energy gaps in the electronic structure spectrum, which also affects the majority of the observed Dirac-like states. Our study constitutes the first experimental observation of the electronic structure of MCo$_2$Al$_9$ and proposes these systems as hosts of Dirac-like physics with intrinsic spin-orbit coupling. The latter effect suggests MCo$_2$Al$_9$ as a future platform for investigating the emergence of non-trivial topology.

\end{abstract}

\maketitle

An intermetallic is a conducting alloy comprised of two or more distinct metals, the crystal structure of which differs from that of the constituents. They have attracted considerable interest because of their ability to feature magnetic orders \cite{Zhang_2019b,Kamishima_2000,Kabir_2022}, superconducting \cite{Cava_1994,Klimczuk_2012,Su_2022}, and half-metal behaviour \cite{Kono_2020,Gao_2022}. Their magnetic and electronic properties are generally attributed to the mixture of metallic, covalent, and ionic atomic bonding. Moreover, their structural complexity makes them a tantalizing platform for chemical substitution, providing the opportunity to tune their properties, often resulting into isostructural series with markedly different electronic and magnetic behaviour. One notable example is given by the Heusler alloys \cite{Heusler_1912,Manna_2018}, for which several magnetic orders have been demonstrated \cite{Belopolski_2019,Finley_2019,Halder_2011}, making them at the top of the list for the possible candidates in spintronics applications \cite{He_2020,Guillemard_2019,Kabanov_2021}.\\

\begin{figure*}
\centering
\includegraphics[width=0.9\textwidth]{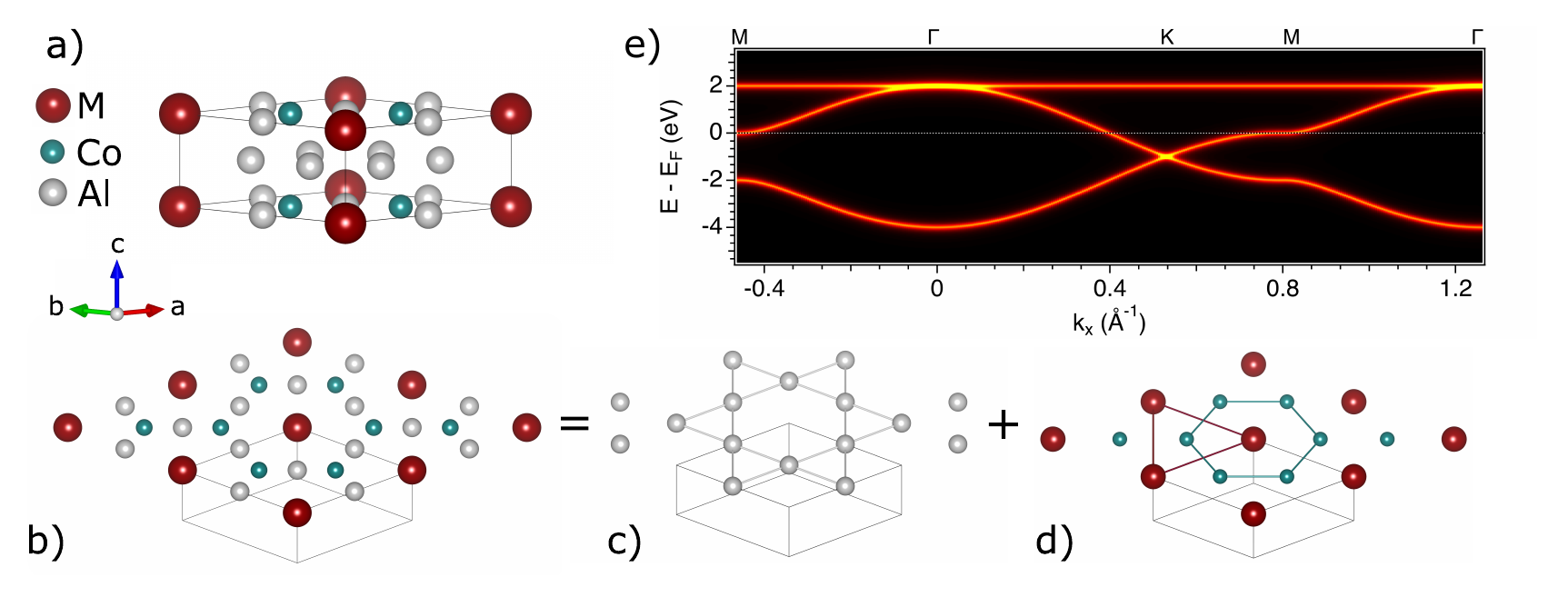}
\caption{MCo$_2$Al$_9$ crystal structure \cite{SCA_CIF} (a) side view and (b) top view showing the atomic arrangement within a single unit cell. The atoms arrange themselves in a complex planar geometry which could be thought as the combination of (c) Al kagome, (d) Co honeycomb and M triangular planes. (e) Example of electronic structure expected for a kagome symmetry obtained by single-orbital first neighbours tight-binding calculations. Due to this particular metric, one can expect the electronic structure to exhibit Dirac-like dispersions and flat bands.}
\label{fig1}
\end{figure*}

MCo$_2$Al$_9$ (M: Ba, Sr) is a new class of intermetallics, isostructural to the sister compound BaFe$_2$Al$_9$, which exhibits the so-called 'catastrophic' charge density wave transition at 100 K, with subsequent shattering of the crystal \cite{Meier_2021}. BaFe$_2$Al$_9$ has also been predicted to host a rich electronic structure (both above and below the charge density wave transition) with Dirac-like dispersions and flat bands, somewhat reminiscent of the electronic structure of kagome and honeycomb systems \cite{Kang_2020a, Okamoto_2022, Han_2021, Polini_2013} (See Fig.\ref{fig1} as an example of how these features can arise from the crystal structure). Motivated by this observation, here, we study the electronic properties of the intermetallic family MCo$_2$Al$_9$, and we demonstrate the existence of Dirac-like bands, some of which are separated by an energy gap due to the spin-orbit coupling (SOC). With this work, we shed light on the electronic properties of MCo$_2$Al$_9$, relating these to hidden structural metrics in the systems' unit cell, and we ultimately propose these systems as possible hosts of SOC-derived non-trivial topological properties \cite{Zhang_2019, Choudhary_2019, Yin_2019, Zhang_2013}.\\

Single crystals MCo$_2$Al$_9$, with M being Sr and Ba, were grown by using the self-flux method described in Ref. \cite{Ryzynska_2020,Canfield_2016}. The powder X-Ray diffraction analysis showed that both compounds crystallize into an hexagonal crystal structure belonging to the \textit{P6/mmm} space group \cite{Turban_1975}. Contrary to BaFe$_2$Al$_9$ \cite{Meier_2021}, both SrCo$_2$Al$_9$ (SCA) and BaCo$_2$Al$_9$ (BCA) do not show any phase transition down to $1.8$~K \cite{Ryzynska_2020}. The crystals were pre-oriented \emph{ex-situ} along high symmetry directions and then cleaved perpendicularly to the c-axis in ultrahigh vacuum (UHV) at the base pressure of $5 \times 10^{-11}$~mbar and at $77$~K temperature, which was kept constant throughout the data acquisition. The metallic bonds of these systems result in a very high hardness, which makes the samples very difficult to break along the c-axis. Nevertheless, we were able to apply sufficient strength to the sample to cleave it in the aforementioned conditions. We noticed that both samples showed a similar and significant quality degradation time of about $12$~h, thus the angle-resolved photoelectron spectroscopy (ARPES) measurements were performed within that time-range. Such a fast degradation time, is consistent with a high chemical reactivity of materials containing alkali-earth metals, such as both BaCo$_2$Al$_9$ and SrCo$_2$Al$_9$. After the cleavage, all samples showed a uniform surface, with no spectroscopic variation at several position of the beam spot (50$\times$100~$\mu$m$^2$). This was also checked for multiple crystals, yielding always the same result, thus with no evidence for a termination-dependent electronic structure and no change in the spectral weight across the surface.\\

\begin{figure*}
\centering
\includegraphics[width=.8\textwidth]{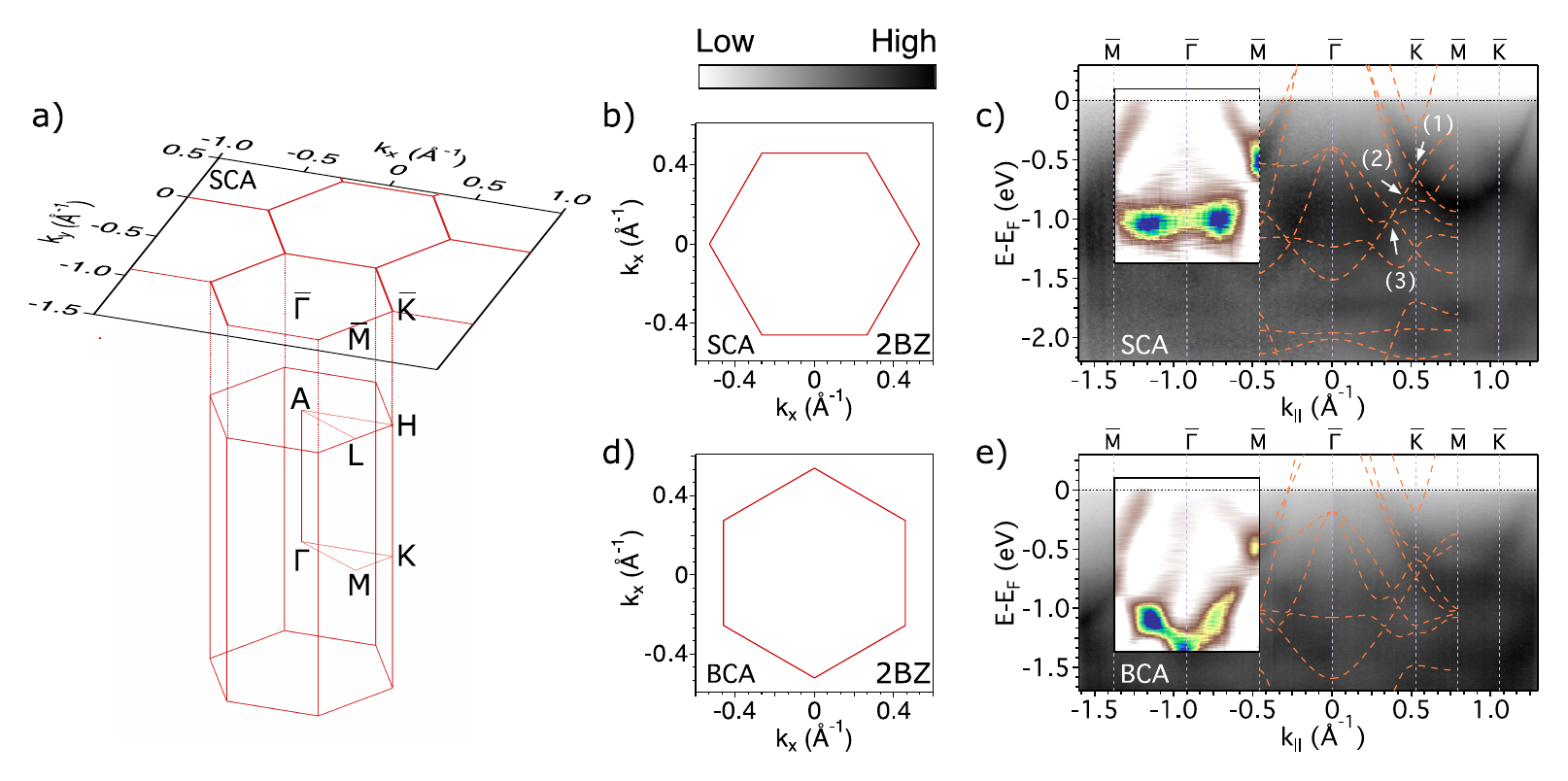}
\caption{(a) BZ and projected surface component along with the Fermi surface of SrCo$_2$Al$_9$ collected at $70$~eV and with linear horizontal polarization. Strong matrix element effects hinders the detection of any photoemitted signal from the first BZ, therefore we carried out the ARPES characterisation in the second zone. (b) second BZ collected for SrCo$_2$Al$_9$ and (c) energy-momentum spectra along the high-symmetry directions indicated. Dashed orange lines superimposed to the ARPES data are the DFT calculated bands (with SOC). Labelled arrows mark the position of the Dirac crossings discussed in the text. (d) second BZ collected for BaCo$_2$Al$_9$ and (e) energy-momentum spectra and DFT-calculated band dispersions (with SOC, dashed lines) along the high-symmetry directions indicated. Insets in c) and e) report the 2D-curvature to show the additional spectral weight close to the Fermi level for BaCo$_2$Al$_9$.}
\label{fig2}
\end{figure*}

The ARPES measurements were performed with different photon energies, and the electronic structure exhibits strongly varying matrix elements as well as strong changes over the photon energy range we used ($40-80$~eV, see supplementary Fig. S1 and S2). The latter indicates the strong three-dimensional character of the electronic structure observed, which is also revealed by the DFT calculations along the out of plane direction (See supplementary Fig. S3 and S4). Such a dispersion results in a broadening of the electronic structure which is intrinsic to the system and generally referred to as kz-broadening. The spectral intensity was prominent at $70$~eV, thus we used this to better visualise the electronic structure of the MCo$_2$Al$_9$ compounds. The Fermi surface of MCo$_2$Al$_9$ are shown in Fig.\ref{fig2} and display a nearly circular hole-like pocket which occupies a large portion of the Brillouin zone (BZ). We estimated a surface-projected carrier density of $\sim0.77$~e$^-$/cell (see methods and supplementary Fig. S1a). Interestingly, the matrix elements create a destructive interference which suppresses almost entirely the ARPES intensity within the first BZ. This can be observed, for example, in Fig.\ref{fig2}a, where the Fermi surface of SrCo$_2$Al$_9$, centred in $(k_x,k_y)=(0,0)$, has no intensity. Interference patterns like the one measured for these compounds are generally a consequence of the matching between initial and final states during the photoemission process. For details on these matrix element effects see refs. \cite{Boschini_2020,Mazzola_2013,Mazzola_2017,Mulazzi_2009,Day_2019}, we report the specific experimental geometry in supplementary, Fig. S1c)). At first glance, the Fermi surface of SrCo$_2$Al$_9$ (Fig.\ref{fig2}b) and BaCo$_2$Al$_9$ (Fig.\ref{fig2}d) appear very similar in shape and size. However, BaCo$_2$Al$_9$ exhibits additional spectral features located at the BZ centre (Fig.\ref{fig2}d). Such an intensity, which is instead not detected in the Fermi surface of SrCo$_2$Al$_9$ (Fig.\ref{fig2}b), can be attributed to the presence of a more p-doped hole-like band in the Ba-based compound, the intensity of which, likely helped by the large broadening of the electronic structure, spills up to the Fermi level. This can be also seen in the DFT (Fig.\ref{fig2}c,e) and highlights one of the most prized properties of isostructural intermetallics: The opportunity of sharing general common features in their electronic, thermal, and mechanical properties, but at the same time, tuning such properties through chemical substitution \cite{Continenza_1992, Meng_2003, Continenza_1992b}.\\

To better understand the origin of the observed spectral weight, we measured the energy-momentum spectra by ARPES for both compounds and we compared them to the DFT calculations. Overall, the DFT calculations (orange lines in Fig.\ref{fig2}c,e) track the experimental bands satisfactorily. Such bands appear broad and this is probably due to the strong chemical bonds between the atoms, which in the out of plane direction can give rise to an extended $k_z$-broadening \cite{Ryzynska_2020}. A strong three-dimensional electronic structure broadening, which is also common to other intermetallics \cite{Chen_2018}, might certainly contribute to such large experimental linewidths. Plus, the three-dimensional character of these systems is also demonstrated by the strong variation along the $k_z$ direction obtained from DFT (See along $\Gamma$-A in Supplementary Fig. S3 and S4). To help the data visualisation, we plotted the curvature \cite{Zhang_2011} signals along the M-$\Gamma$-M direction (See insets in Fig.\ref{fig2}c,e). These plots readily reveal the existence of the extra intensity observed in BaCo$_2$Al$_9$ and centred in $\Gamma$. By comparing the data with the DFT calculations, we ascribe such a feature to the hole-like bands dispersing with maximum at $\Gamma$. Notably, the top is closer to the Fermi level for the Ba-derived compound than the Sr one, consistent with the additional spectral intensity observed in the BaCo$_2$Al$_9$ Fermi surface with respect to SrCo$_2$Al$_9$ (Fig\ref{fig2}c,e). The observed energy shift is large and estimated to be $\sim250$~meV. We stress that despite the observed hole-like band is located below the Fermi level, the large tails of spectral intensity extend up to it, thus being still observable in the Fermi surface of Fig.\ref{fig2}d.\\

\begin{figure*}
\centering
\includegraphics[width=.8\textwidth]{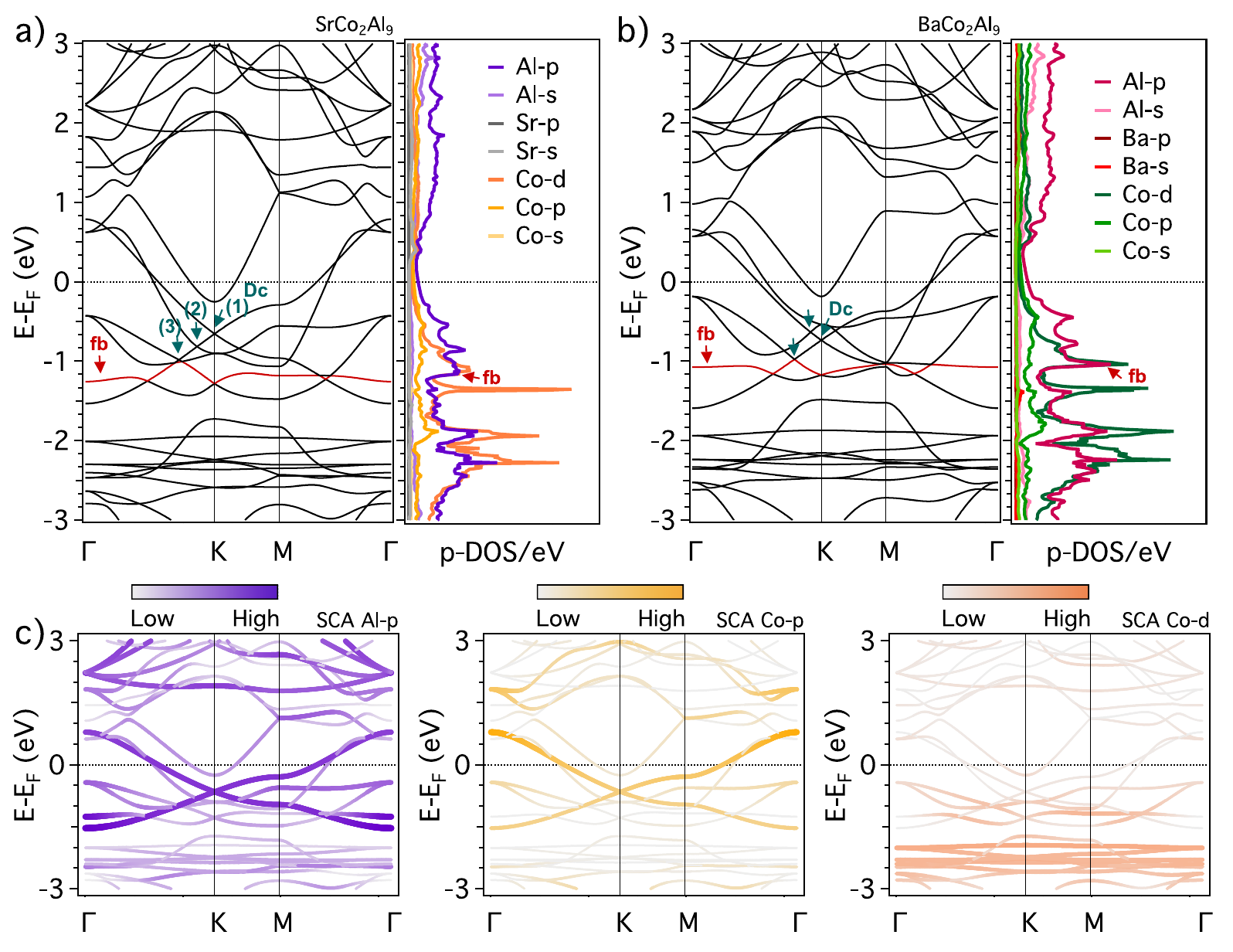}
\caption{\emph{Ab-initio} bulk electronic band dispersion extracted along high symmetry directions in the $\Gamma$-K-M-$\Gamma$ plane and orbital-projected partial DOS for (a) SrCo$_2$Al$_9$ and (b) BaCo$_2$Al$_9$. Labelled arrows (1)-(3) pin the Dirac-like crossings ('Dc') while the flat band ('fb') linked to the hidden Al-kagome lattice is highlighted in red and marked by the red arrows both in the band dispersion and in the p-DOS. (c) Orbital character of the SrCo$_2$Al$_9$ band structure along the same high symmetry path shows that Al-p, Co-p and Co-d orbitals bring the highest contribute to the states near E$_F$}
\label{fig3}
\end{figure*}

\begin{figure*}
\centering
\includegraphics[width=0.8\textwidth]{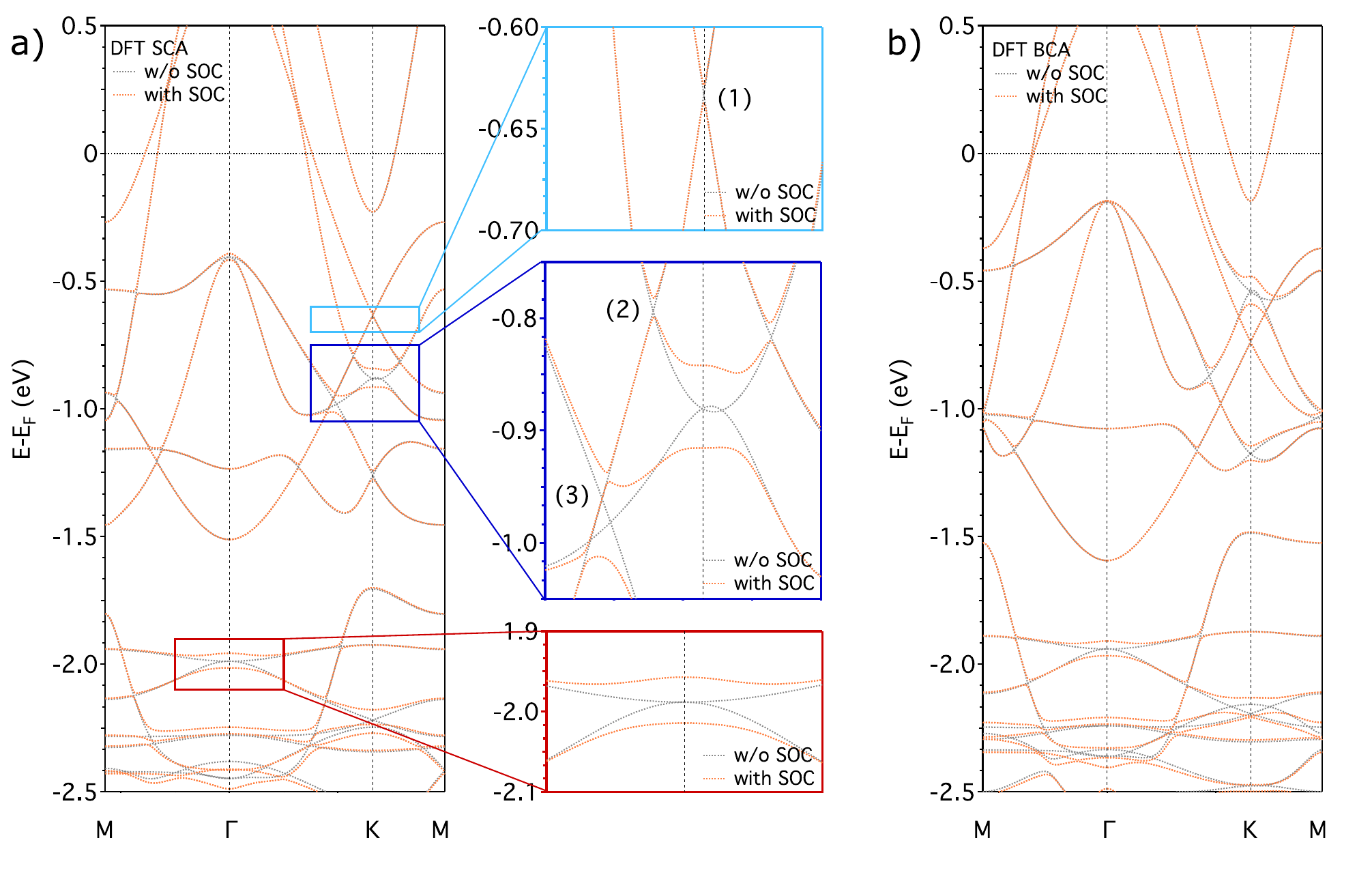}
\caption{Effects of SOC on the electronic band structure of a) SrCo$_2$Al$_9$ and b) BaCo$_2$Al$_9$. Zoom-in of the coloured boxes in a) shows details of the anti-crossing opening gaps induced by SOC at K and $\Gamma$ points. Labels (1)-(3) mark the positions of the Dirac crossings discussed above.}
\label{fig4}
\end{figure*}

To better capture the details of the electronic structure of both MCo$_2$Al$_9$, we show the DFT (without SOC) along with the calculated projected electron density of states (p-DOS, See Fig.\ref{fig3}). First of all, we notice that in both compounds, the main orbital character projected on the eigenstates is provided by the Al and the Co atoms, while the M metals seem to only contribute (in a less prominent way) to the electronic structure far away from the Fermi level (energies larger than $-2$~eV). The M metals, despite responsible for energetic shifts, as discussed above, do not appear to contribute significantly to the near-Fermi electronic structure and fermiology. In the p-DOS (See Fig.\ref{fig3}a,b), several peaks with Co d-orbital character contribute to the main signal at binding energy smaller than $-1$~eV and such peaks originate from nearly dispersionless bands. Such Co-derived localised states are not primarily linked to the flat band arising from the Al kagome motif, but rather to the strong correlations experienced by electrons in the d-shell. Incidentally, we note that the sharp and intense peaks at E$_B\sim-1.1$~eV and at E$_B\sim-1.35$~eV in the Co-d p-DOS are due to highly localised Co-d bands which lie at the edge of the Brillouin zone in the A-L-H plane (see supplementary Fig. S3, S4 where the whole orbital-projected bulk bands are provided) rather than to the kagome's flat band highlighted in red in Fig.\ref{fig3}a ('fb'). The latter possesses indeed essentially Al-p orbital character, with marginal contribution from Co-d hybridisation, as shown in Fig.\ref{fig3}c. The presence of flat bands in these materials is somewhat different from what reported for kagome metals. In the latter, in fact, they arise from a phase-cancellation process: In kagome, the eigenstates of the flat band have opposite sign at different sublattices, as discussed in Ref.\cite{Li_2021, Zhi_2018, Kang_2020a, Kang_2020b}. Therefore, any electron hopping outside the hexagonal part of the kagome lattice is cancelled out by a quantum destructive interference. This cancellation gives rise (within a first-neighbours tight binding model, See Fig.\ref{fig1}) to a perfect localization of electrons within the BZ. In pure kagome, both the effects of hybridization with the next-neighbouring atoms and SOC can gap these states. However, their major contribution will still be given by the geometrical kagome pattern. Here, for our systems, while the Al atomic arrangement is reminiscent of a kagome lattice, there is a similar orbital contribution to such states given by the Co atoms, which develop with an honeycomb motif. The strong hybridization and overlap between the Al and Co is expected to enable bands with dispersion significantly different from the expected textbook behaviour of a pure kagome mesh. However, our calculations indicate that the flat bands (highlighted with red colour in Fig.\ref{fig3}a - energy $\sim-1.2$~eV - and in Fig.\ref{fig3}b - energy $\sim-1$~eV - for SrCo$_2$Al$_9$ and BaCo$_2$Al$_9$, respectively) are a genuine effect of the Al atomic arrangement, since they are primarily Al-derived (Fig.\ref{fig3}c). Evidence for such flat states is also visible in the ARPES spectra of Fig.\ref{fig2}, however they are better highlighted when LV polarisation was used as this suppressed the intensity from the highly dispersive bands (see Fig.S2 in supplementary where we report such spectra together with their curvature).\\

The intrinsic honeycomb and kagome motifs created by the Co and Al respectively in both compounds is evident from a pronounced Dirac-like dispersion with crossing point at K. Such a feature is clearly visible both in the experiment and in the DFT calculations, e.g. see arrow (1) in Figs.\ref{fig2}c,\ref{fig3}a (E$_B\sim-0.6$~eV) and also Fig.\ref{fig3}c, indicating the hidden symmetries nestled in the crystal structure of both intermetallics. This band crossing is present for both compounds and has van-Hove singularity at the M point approximately $300$~meV($400$~meV) below the Fermi level for SrCo$_2$Al$_9$(BaCo$_2$Al$_9$). Due to strong matrix element effects its presence in Fig.\ref{fig2} is concealed by the intense highly dispersive states along K-M (for better visualisation see supplementary Fig. S5 where we probe the van-Hove singularity in different geometries to better highlight it). According to our orbital projected DFT, similarly to graphene, such a Dirac dispersion appear to derive from p-type orbitals, in this case with mixed Al-Co character. We notice also that the same bands which develop into a Dirac-like dispersion are the same that in the experiment give rise to the largest circular Fermi surface as in Fig.\ref{fig2}. We notice that such bands, in proximity of the Fermi level are very close to each other, thus giving rise, within our resolution, to a single ring rather than the two theoretically expected. If on the one hand the MCo$_2$Al$_9$ bears several similarities with graphene, we notice remarkable differences: in graphene, the energy difference between the Dirac crossing and the van-Hove singularity is large, i.e. approximately $3$~eV \cite{Hellsing_2018, Bisti_2015}. Here, it is instead one order of magnitude smaller, i.e. approximately $0.3$~eV. This strikingly different behaviour could be exploited, by filling or depopulating the singularity point thanks to the ease of tuneability offered by this family of materials. Once these states are brought to the Fermi level, they could lead to collective phase transition. A notable example is the BaFe$_2$Al$_9$ where the Al-derived van-Hove singularity and Dirac cones cross the Fermi level and BaFe$_2$Al$_9$ undergoes a charge density wave transition so strong that the crystal shatters \cite{Meier_2021}. Finally, the lower branch of the Dirac cone intersects other highly dispersive bands along $\Gamma$-K, see for example E$_B\sim-0.8$~eV (green arrow (2)) and E$_B\sim-1$~eV (green arrow (3)) in Fig.\ref{fig3}a. These crossings bands are also characterised by a strong intermixing of Al-Co orbitals, with the only difference that this time are the Co d-orbitals that hybridises with the Al-p character.\\

SOC further mixes the orbital character of the bands, introducing energy gaps in the MCo$_2$Al$_9$ band structure (See Fig.\ref{fig4} for bands with and without-SOC, orange and grey, respectively). According to the DFT, the effect of the SOC is stronger in the bands displaying a Co-d character and manifests by gapping Dirac-like states at both K and $\Gamma$ points. Specifically, SrCo$_2$Al$_9$ displays such energy gaps of $\sim70$~meV at K point (blue framed zoom) and of $\sim60$~meV at $\Gamma$ (red frame). In the kagome metals such as the recently discovered XV$_6$Sn$_6$ family (X: Gd, Y, Tb) \cite{Pokharel_2021,Pokharel_2022,Rosenberg_2022}, spin-orbit induced gaps in the electronic structure have attracted attention because they are expected to host a finite contribution to the spin-Berry curvature. This has been recently proved by Di Sante et al. in ref.\cite{DiSante_2023} and it applies also to other systems as partially discussed in ref.\cite{Mazzola_2023} for TaCoTe$_2$. Our system, which intrinsically hosts this lattice is not expected to behave differently (and with the additional ease of intermetallics to tune such properties even further by chemical substitution), thus our work invites further study of this topological aspect.\\

Summarising; the combination of ARPES and DFT allows us to discover the presence of several Dirac-like dispersions in the MCo$_2$Al$_9$ intermetallic family, hidden in the honeycomb/kagome sub-structures. Orbital-projected DFT consistently pinpoints a strong Al-Co hybridisation across the whole Dirac-like band structure. Finally, our DFT reveal the existence of sizeable SOC-induced gaps in most of the crossing. We therefore establish the benchmarks of the electronic properties of these compounds and pave the way for further studies of SOC-derived phenomena and the development of non-trivial topology in the series. 

\section*{Acknowledgements}
This work has been performed in the framework of the Nanoscience Foundry and Fine Analysis (NFFA-MIUR Italy Progetti Internazionali) facility (www.Trieste.NFFA.eu). CB and PDCK gratefully acknowledge support from The Leverhulme Trust via Grant No. RL-2016-006. F.M. greatly acknowledges the SoE action of pnrr, number SOE\_0000068. K. S. T and S. P. acknowledge funding from the European Research Council (ERC) under the European Union's Horizon 2020 research and innovation program Grant No. 773122 (LIMA). K. S. T. is a Villum Investigator supported by VILLUM FONDEN (grant no. 37789).

\bibliography{references.bib}
\end{document}

\title{The electronic structure of intertwined kagome, honeycomb, and triangular sublattices of the intermetallics MCo$_2$Al$_9$}

\author{Chiara Bigi}\email{chiara.bigi@synchrotron-soleil.fr}
\affiliation{SUPA, School of Physics and Astronomy, University of St Andrews, St Andrews KY16 9SS, UK}\affiliation{Synchrotron SOLEIL, F-91190 Saint-Aubin, France}

\author{Sahar Pakdel}
\affiliation{CAMD, Department of Physics, Technical University of Denmark, 2800 Kgs. Lyngby, Denmark}

\author{Micha{\l} J. Winiarski}
\affiliation{Faculty of Applied Physics and Mathematics, Advanced Materials Centre, Gdansk University of Technology, Narutowicza 11/12, 80-233, Gdansk, Poland}

\author{Pasquale Orgiani}
\affiliation{Istituto Officina dei Materiali (IOM)-CNR, Laboratorio TASC, Area Science Park, S.S.14, Km 163.5, 34149 Trieste, Italy}

\author{Ivana Vobornik}
\affiliation{Istituto Officina dei Materiali (IOM)-CNR, Laboratorio TASC, Area Science Park, S.S.14, Km 163.5, 34149 Trieste, Italy}

\author{Jun Fujii}
\affiliation{Istituto Officina dei Materiali (IOM)-CNR, Laboratorio TASC, Area Science Park, S.S.14, Km 163.5, 34149 Trieste, Italy}

\author{Giorgio Rossi}
\affiliation{Istituto Officina dei Materiali (IOM)-CNR, Laboratorio TASC, Area Science Park, S.S.14, Km 163.5, 34149 Trieste, Italy}\affiliation{Department of Physics, University of Milano, 20133 Milano, Italy}

\author{Vincent Polewczyk}
\affiliation{Istituto Officina dei Materiali (IOM)-CNR, Laboratorio TASC, Area Science Park, S.S.14, Km 163.5, 34149 Trieste, Italy}

\author{Phil D.C. King}
\affiliation{SUPA, School of Physics and Astronomy, University of St Andrews, St Andrews KY16 9SS, UK}

\author{Giancarlo Panaccione}
\affiliation{Istituto Officina dei Materiali (IOM)-CNR, Laboratorio TASC, Area Science Park, S.S.14, Km 163.5, 34149 Trieste, Italy}

\author{Tomasz Klimczuk} 
\affiliation{Faculty of Applied Physics and Mathematics, Advanced Materials Centre, Gdansk University of Technology, Narutowicza 11/12, 80-233, Gdansk, Poland}

\author{Kristian Sommer Thygesen}\email{thygesen@fysik.dtu.dk}
\affiliation{CAMD, Department of Physics, Technical University of Denmark, 2800 Kgs. Lyngby, Denmark}

\author{Federico Mazzola}\email{federico.mazzola@unive.it}
\affiliation{Department of Molecular Sciences and Nanosystems, Ca’ Foscari University of Venice, 30172 Venice, Italy}\affiliation{Istituto Officina dei Materiali (IOM)-CNR, Laboratorio TASC, Area Science Park, S.S.14, Km 163.5, 34149 Trieste, Italy}

\date{\today}
\maketitle
%\newpage
\section*{Methods}
{\bf Tight binding model.} The electronic structure for Al s-orbitals arranged in a kagome lattice in Fig.1e was obtained for on-site energy E$_0=0$~eV and hopping $\sigma$ bond V$_ss=-1$~eV. The nearest neighbour hopping distance was set equal to the Al-Al kagome net in the $ab$-plane of MCo$_2$Al$_9$ system (a=b=$7.9$~\AA). An impurity scattering broadening term Im($\Sigma$)~=~$100$~meV was introduced to simulate the finite lifetime of a realistic spectral function.\\

{\bf Experimental details.} Single crystals of SrCo$_2$Al$_9$ and BaCo$_2$Al$_9$ were grown employing the self-flux technique \cite{Canfield_1992} using alumina crucibles and a frit-disc \cite{Canfield_2016}. Sr ($99.95\%$ pure, Onyxmet) or Ba dendrites ($99.95\%$, Onyxmet), Co scraps ($99.9\%+$, Alfa Aesar) and Al slug ($99.999\%$, Onyxmet) were put in an alumina crucible at the atomic ratio of $2:3:60$ (ca. $1.2$~g total for each sample) and sealed in a quartz tube, evacuated and backfilled with Ar to dilute the Al vapor attacking the tube walls. The ampoules were placed in a box furnace, heated to $1120^{\circ}$~C, held for $6$~hours and then slowly cooled ($2^{\circ}$~C/h) to $900^{\circ}$~C (SrCo$_2$Al$_9$) or $845^{\circ}$C (BaCo$_2$Al$_9$). The temperature program and sample compositions were adjusted (based on recent crystal growth reports \cite{Ryzynska_2020,Meier_2021}) to yield single crystals free of SrAl$_4$/BaAl$_4$ contamination and of sufficient size for ARPES measurements. At the final temperature, each ampoule was centrifugated to remove the excess Al-rich flux. The growth process yields numerous hexagonal needle-like single crystals, the largest being ca. $1$~mm wide and ca. $5$~mm long. The remaining droplets of solidified Al were etched from the crystal surfaces using ca. $0.1$~M NaOH solution in an ultrasonic bath. Selected single crystals were fine-ground and examined using a powder x-ray diffractometer (Bruker D2 Phaser instrument, CuK$\alpha$ source, LynxEye XE-T detector) to confirm the phase composition. Obtained powder patterns were consistent with previous reports \cite{Ryzynska_2020,Turban_1975}. The ARPES measurements were carried out at the synchrotron radiation source ELETTRA by using the DA30 hemispherical analyser of the APE-LE end-station (Trieste). The energy and momentum resolutions were better than $12$~meV and $0.018$~\AA$^{-1}$, respectively. The sample temperature was kept constant at $77$~K throughout the experiment. Radial momentum distribution curves (MDC) were extracted every $5$ degrees over $180^{\circ}$ range from the SrCo$_2$Al$_9$ Fermi surface and fitted to Lorentzian peak to extract the projected electron carrier density. A factor two was introduced to account for the two bands forming the Fermi contour as resulting from DFT calculations.\\

{\bf DFT.} First-principles calculations are performed using density functional theory (DFT) \cite{PhysRev.140.A1133} within the projector-augmented wave (PAW) formalism \cite{PhysRevB.50.17953} as implemented in the GPAW code \cite{Enkovaara_2010, PhysRevB.71.035109}. All the calculations were performed using the Atomic Simulation Recipes (ASR) framework \cite{GJERDING2021110731} in combination with the atomic simulation environment (ASE) \cite{Larsen_2017}. The exchange correlation potentials were described through the Perdew-Burke-Ernzerhof (PBE) functional within the generalized gradient approximation (GGA) formalism \cite{perdew}. A plane-wave cutoff energy of 800 eV is employed for all calculations and the width of the Fermi-Dirac smearing is kept at 50 meV. For geometry optimization, 6/$\AA^{-1}$ kmesh density is used and forces are converged until 0.01 eV/$\AA^{-1}$. After lattice relaxation, the electron density is determined self-consistently on a uniform k-point grid of density 12.0/$\AA^{-1}$. From this density, the PBE band structure is computed non-self consistently at 400 k-points distributed along the band path. Spin-orbit coupling (SOC) is added non-self consistently.\\

\clearpage
\begin{figure*}
\centering
\includegraphics[width=0.7\textwidth]{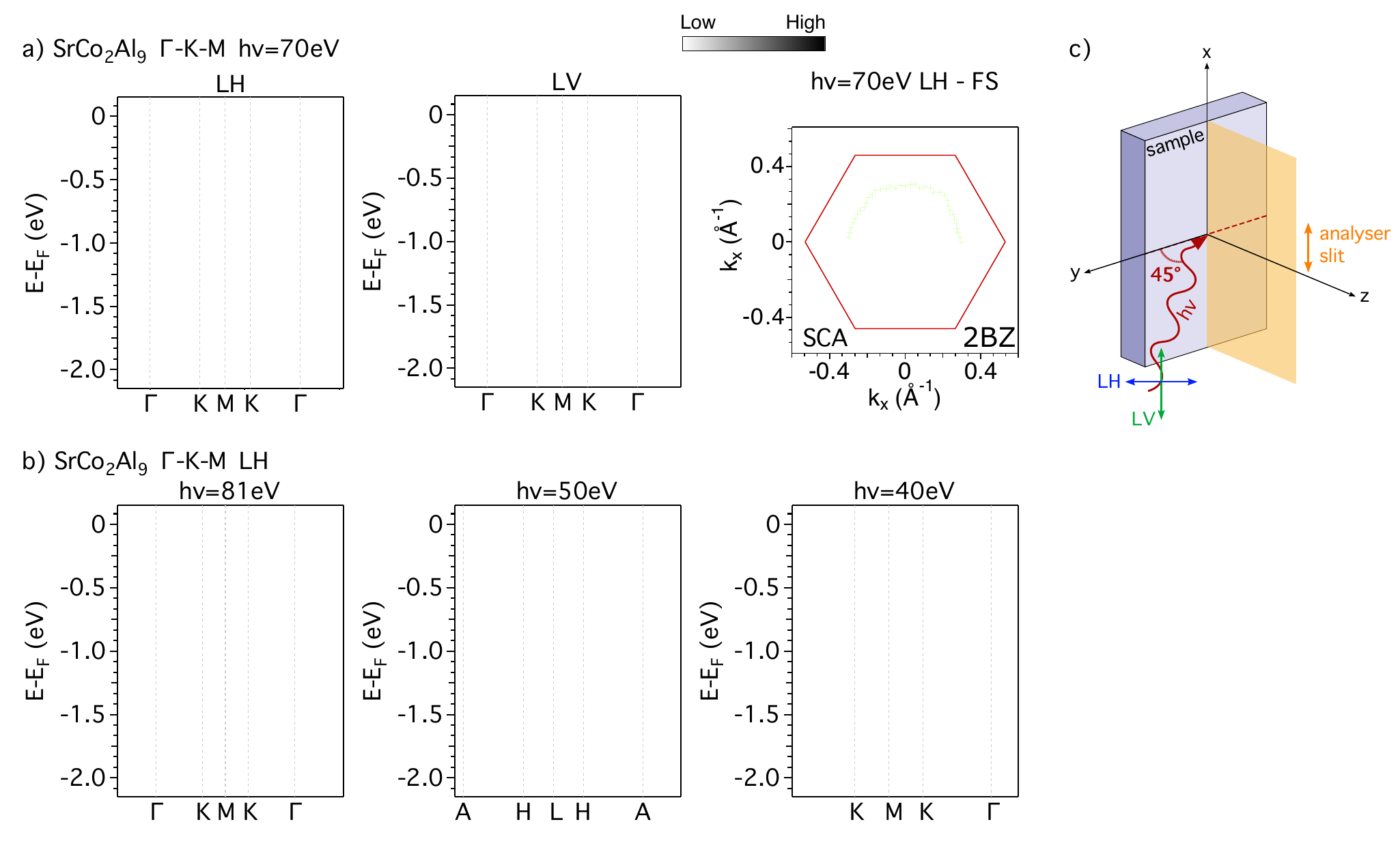}
\caption{(a) SrCo$_2$Al$_9$ collected along the $\Gamma$-K-M direction at $70$~eV photon energy with both linear horizontal (LH) and linear vertical (LV) polarisation showing the strong matrix elements' dependence of the Al-p derived bands. The green markers on the second BZ Fermi surface of SrCo$_2$Al$_9$ indicate the fitting of the Fermi contour from which the projected carrier density of $\sim0.77$~e$^-$/cell was estimated. (b) SrCo$_2$Al$_9$ measured along the $\Gamma$-K-M direction at other photon energies (i.e. $81$~eV, $50$~eV and $40$~eV) and LH polarisation to show the strong matrix element effects. (c) Sketch of our experimental geometry, where the photons impinge the sample surface with a $45^{\circ}$~ angle. Linear vertical (LV, green arrow) polarization will be completely within the plane of the sample and parallel to the analyser slit, in particular along the x direction as in the figure. On the contrary, linear horizontal (LH, blue arrow) polarization will be split 50\% in plane (along y) and 50\% out of plane (along z).  Simplistically, one can imagine that LH will couple also to orbital owing an out of plane dispersion. For details on matrix elements, see Ref. \cite{Boschini_2020,Mazzola_2013,Mazzola_2017}}
\label{S1}
\end{figure*}

\begin{figure*}
\centering
\includegraphics[width=0.7\textwidth]{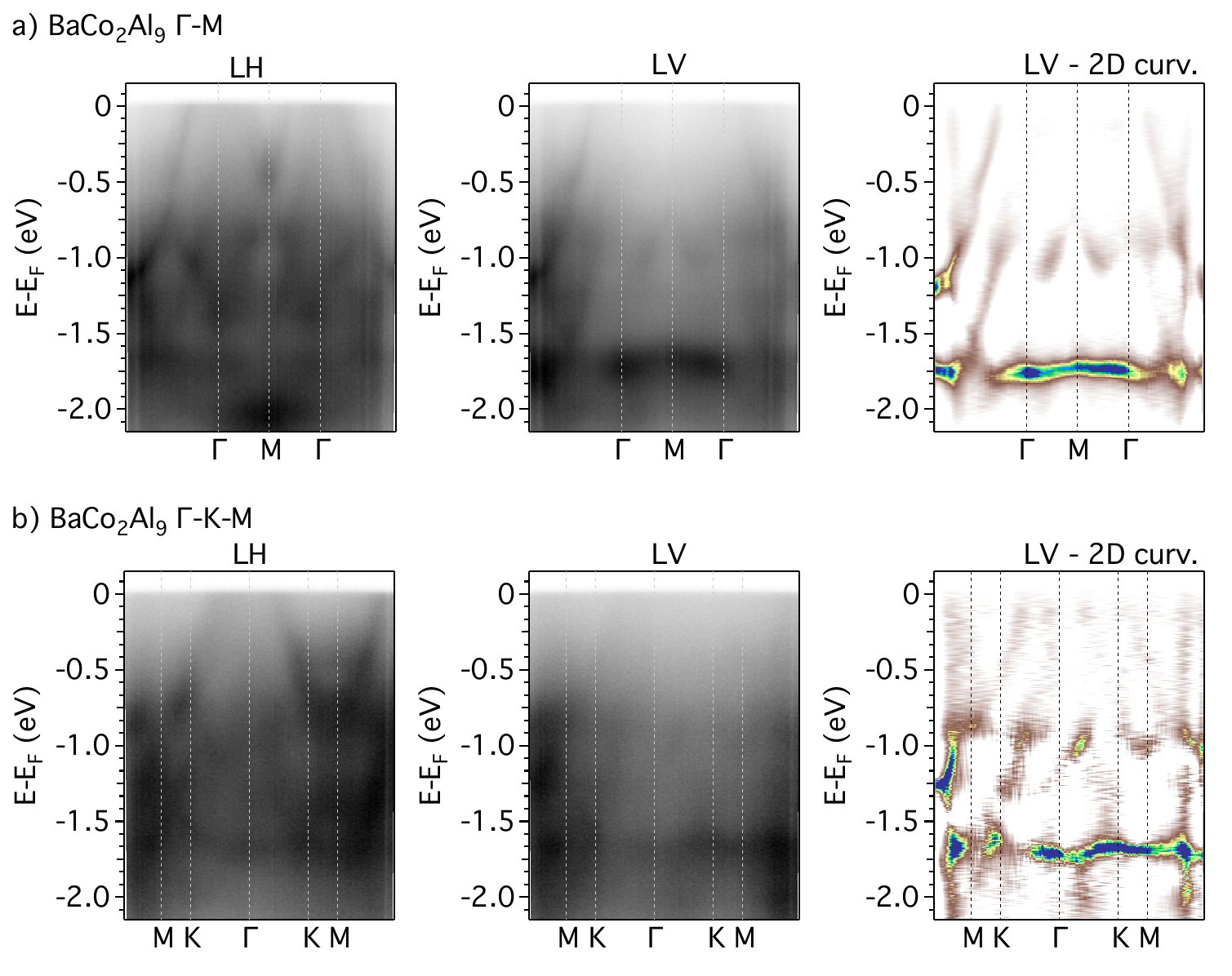}
\caption{BaCo$_2$Al$_9$ collected at $70$~eV photon energy with both LH and LV polarisation along the and along the (a) $\Gamma$-M and (b) $\Gamma$-K-M directions which shows the strong matrix elements' suppression of the photoemitted intensity when LV polarisation is used.}
\label{S2}
\end{figure*}

\begin{figure*}
\centering
\includegraphics[width=0.9\textwidth]{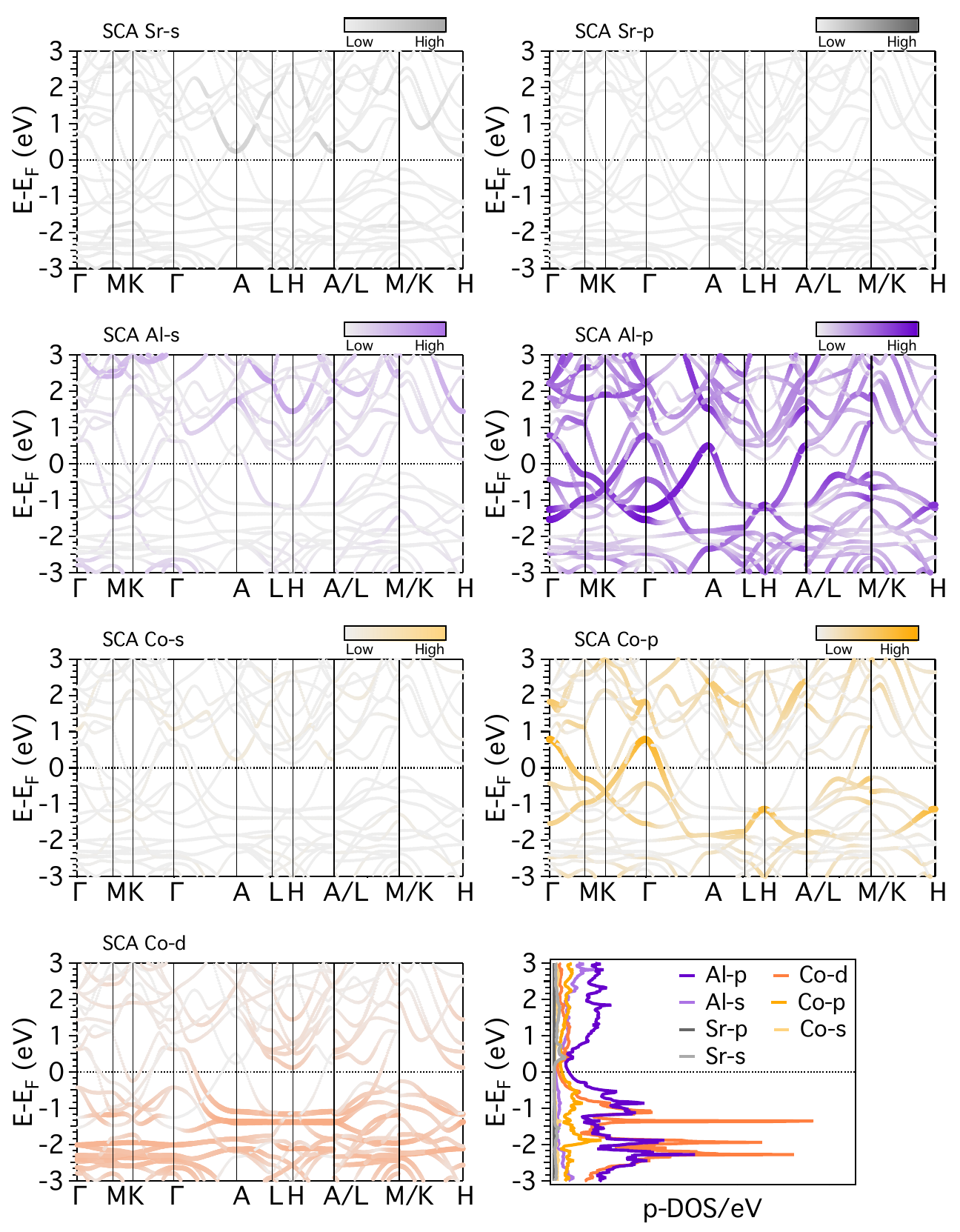}
\caption{Orbital projected band structure and electronic density of states of SrCo$_2$Al$_9$ as obtained from DFT without SOC. Thicker lines in the bands mark bigger weights of the particular atomic orbital to the electronic structure. To compare between different orbital character the intensity scale has been normalised to the Al-p character which has the highest weight in the proximity of the Fermi level. The k-integrated projected electronic density of states is also reported.}
\label{S3}
\end{figure*}

\begin{figure*}
\centering
\includegraphics[width=0.9\textwidth]{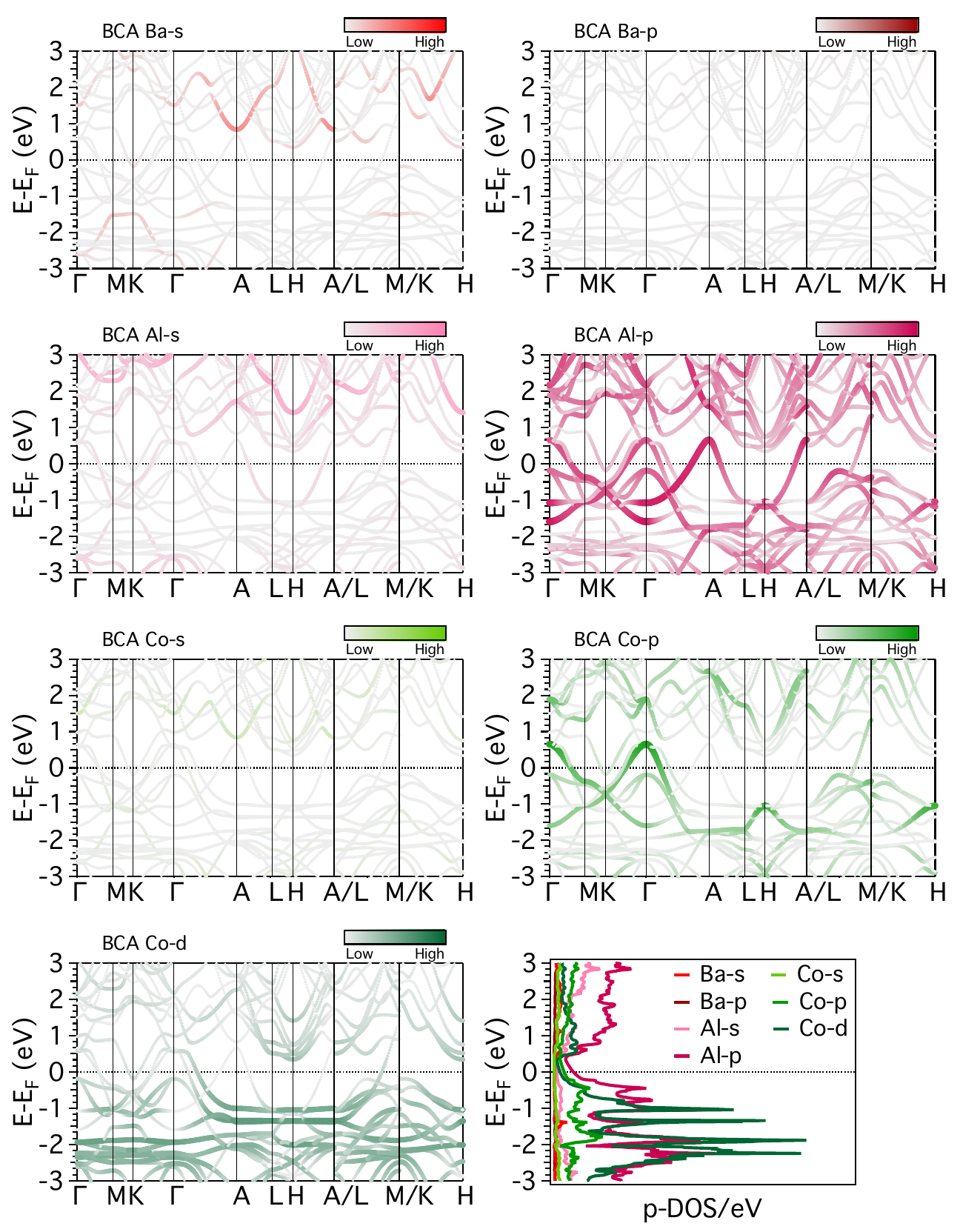}
\caption{Similar to Fig.\ref{S3}, orbital projected band structure and electronic density of states of BaCo$_2$Al$_9$ as obtained from DFT without SOC.}
\label{S4}
\end{figure*}
\clearpage
\begin{figure*}
\centering
\includegraphics[width=0.9\textwidth]{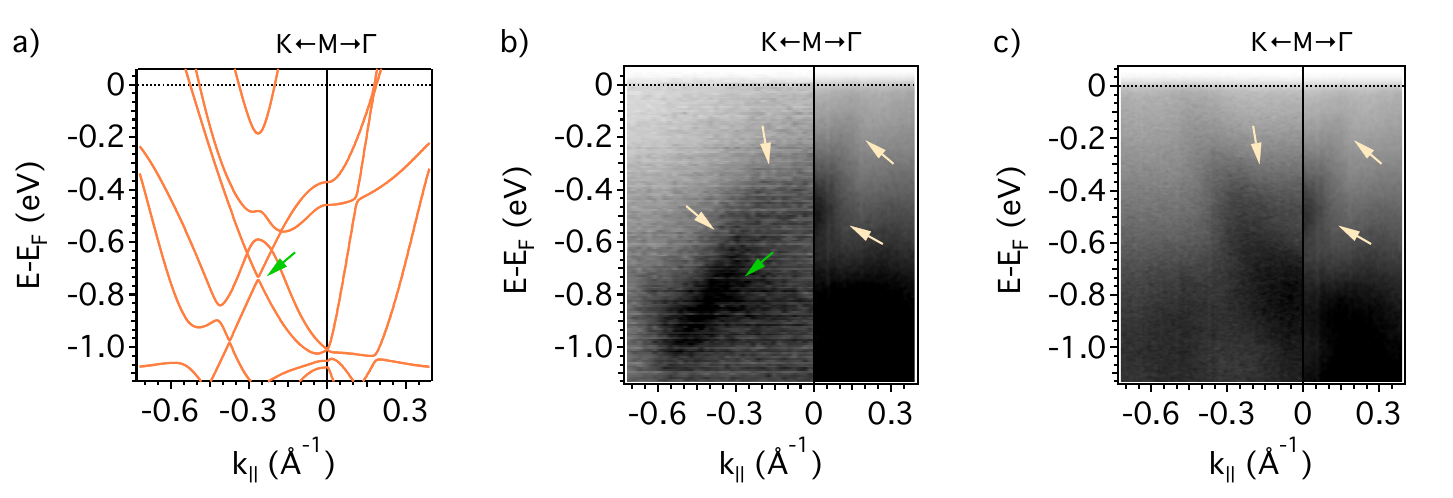}
\caption{The van Hove singularity in BaCo$_2$Al$_9$ along K-M-$\Gamma$ high symmetry directions a) \emph{ab-initio} calculations (with SOC) and b), c) ARPES data extracted along the same high symmetry directions. Due to the strong matrix element effects the van-Hove singularity is concealed by the bands along K-M, to better visualise it, we show the data collected for two different experimental geometries b) K-M orthogonal to the analyser slit (i.e., extracted from Fermi surface map measured on a sample oriented along $\Gamma$-M) and c) K-M along the analyser slit (i.e., different cleave for which the sample was oriented along $\Gamma$-K). Incidentally, we note that the enhanced intensity at $\sim-0.7$~eV is consistent with the position of the predicted Dirac cone (marked with green arrow in panels a) b).}
\label{S5}
\end{figure*}

\bibliography{references.bib}